# ADAPTIVE TYPE-2 FUZZY SECOND ORDER SLIDING MODE CONTROL FOR NONLINEAR UNCERTAIN CHAOTIC SYSTEM

Rim Hendel[1], Farid Khaber[1] and Najib Essounbouli[2]

[1] QUERE Laboratory, Engineering Faculty, University of Setif 1, 19000 Setif, Algeria

[2] CReSTIC of Reims  Champagne-Ardenne University, IUT de Troyes, France

## ABSTRACT

*In this paper, a robust adaptive type-2 fuzzy higher order sliding mode controller is designed to stabilize the unstable periodic orbits of uncertain perturbed chaotic system with internal parameter uncertainties and external disturbances. In Higher Order Sliding Mode Control (HOSMC),the chattering phenomena of the control effort is reduced, by using Super Twisting algorithm. Adaptive interval type-2 fuzzy systems are proposed to approximate the unknown part of uncertain chaotic system and to generate the Super Twisting signals. Based on Lyapunov criterion, adaptation laws are derived and the closed loop system stability is guaranteed. An illustrative example is given to demonstrate the effectiveness of the proposed controller.*

## 1. INTRODUCTION

Chaotic phenomenon is widely observed in several applications such as: medical field, fractal theory, electrical circuits and secure communication [1]. Although, the prominent characteristics of chaotic system is its extreme sensitivity to initial conditions and its unpredictability; it is usually difficult to predict exactly the behavior of the chaotic system. Recently, several researchers have focused on chaos control [2]. Many nonlinear control techniques have been successfully applied on chaos control and synchronization of different dynamical systems [3-5], nonlinear control [6-7], active control and backstepping design [8-10], fuzzy logic and adaptive control [11-12], adaptive fuzzy control [13].

Unfortunately, in the most of the approaches mentioned above the unknown parameters of the chaotic system,the uncertainties,internal and external disturbances, have not been considered, which implies that the robustness has not been investigated. Sliding Mode Control (SMC) is often adopted, due to its inherent advantages of fast dynamic response, guaranteed stability, robustness against matching external disturbances, andinternal parameter variations. Several controllers based on sliding mode control have been proposed for chaos schemes [14-16].

However, it should be noted that the smoothness of a control signal in sliding mode is not easily achievable without loss performance and robustness degradation. A lot of works have been proceeded to solve this problem by using adaptive control [17-18], and intelligent approaches [19-20].

The High Order Sliding Mode Control (HOSMC) has been presented to reduce and (or) remove the chattering phenomenon. Moreover, this technique provides higher accuracy than the standardSMC [21-23]. Higher order sliding modes (HOSM) generalize the basic so-called first order sliding mode idea acting on the higher order time derivatives of sliding function. In the case of second order sliding mode, the sliding set is described as $S = \{s = \dot{s} = 0, \ddot{s} \neq 0\}$, and the control is acting on the second derivative of the switching manifold $s$ [24-25]. A HOSMC has a finite time convergence, which is satisfied when the switching gains in the HOSM control law are selected properly. Nevertheless, the calculation of these gains needs the well knowledge of the system dynamic [21,26].

In this paper, a higher order sliding mode controlcombined with adaptive type-2 fuzzy systems, is proposed to design a robust controller for stabilization of unknown SISO nonlinear chaotic system, working in the presence of uncertainties and external disturbances. The Super Twisting algorithm is implemented to avoid a chattering phenomenon. In the same time, we introduced adaptive type-2 fuzzy systems for model the unknown dynamic of system and simplify the calculation of gains in the second order sliding mode. Their updates are performed using adaptation laws derived from the stability studyin the Lyapunov sense.

The organization of this paper is as follows. In section 2, the problem states and description of the system.The adaptive type-2 fuzzy second order sliding mode control scheme is presented in section III. Simulation example demonstrate the efficiently of the proposed approach in section IV. Finally, section V gives the conclusions of the advocated design methodology.

## 2. DESCRIPTION OF SYSTEM AND PROBLEM FORMULATION

Consider $n$-order uncertain chaotic system which has an affine form:
$$\begin{cases} \dot{x}_i = x_{i+1}, & 1 \leq i \leq n-1, \\ \dot{x}_n = f(\underline{x}, t) + \Delta f(\underline{x}, t) + d(t) + u(t) \end{cases} \tag{1}$$

where $\underline{x} = [x_1(t)\ x_2(t) \dots x_n(t)] \in \Re^n$ is the measurable state vector, $f(\underline{x}, t)$is unknown nonlinear continuous and bounded function, $u(t) \in \Re$is control input of the system, $\Delta f(\underline{x}, t)$and $d(t)$are the uncertainties and external bounded disturbances, respectively,

$$\left| f(\underline{x}, t) \right| < F \quad, \quad \left| \Delta f(\underline{x}, t) \right| \leq \Delta_f \quad, \quad |d(t)| \leq \Delta_d \tag{2}$$
where$F$ , $\Delta_f$ and $\Delta_d$ are positive constants.

The control objective is getting the system to track an $n$- dimensional desired vector $\underline{y}_d(t)$which belong to a class of continuous functions on$[t_0, \infty]$. Let's the tracking error as;

$$\begin{aligned} \underline{e}(t) &= \underline{x}(t) - \underline{y}_d(t) \\ &= [x(t) - y_d(t) \quad \dot{x}(t) - \dot{y}_d(t) \quad \dots \quad x^{(n-1)}(t) - y_d^{(n-1)}(t)\,] \\ &= [e(t) \quad \dot{e}(t) \dots e^{(n-1)}(t)] \end{aligned} \tag{3}$$

Therefore, the dynamic errors of system can be obtained as;
$$\begin{cases} \dot{e}_1 = e_2 \\ \dot{e}_2 = e_3, \\ \vdots \\ \dot{e}_n = f(\underline{x}, t) - y_d^{(n)}(t) + \Delta f(\underline{x}, t) + d(t) + u(t) \end{cases} \tag{4}$$

The control goal considered is that;
$$\lim_{t \to \infty} \left\| \underline{e}(t) \right\| = \lim_{t \to \infty} \left\| \underline{x}(t) - \underline{y}_d(t) \right\| \to 0, \tag{5}$$

## 2.1. Second Order Sliding Mode Control

The basic concept of second order sliding mode control can be interpreted from the following the following second order nonlinear system:

$$\begin{cases} \dot{x}_1 = x_2 , \\ \dot{x}_2 = f(\underline{x},t) + D(\underline{x},t) + u(t), \end{cases} \tag{6}$$

$D(\underline{x},t)$ isthe whole uncertainties indicating the sum of the external disturbances and parameter uncertainties, where $D(\underline{x},t) \le \Delta$ and $\Delta = \Delta_f + \Delta_d$.

The linear sliding manifold is defined as,

$$s(\underline{e},t) = \left(\frac{\partial}{\partial t} + \lambda\right)^{(n-1)} \underline{e} \tag{7}$$

where $\lambda > 0$ is a positiveconstant, The time derivative of $s$ is:

$$\dot{s}(\underline{e},t) = e^{(n)} + \delta_s$$

where $\delta_s = \sum_{k=1}^{n} \frac{(n-1)!}{k!(n-k-1)!} \left(\frac{\partial}{\partial t}\right)^{(n-k-1)} \lambda^k \underline{\dot{e}}$.

By using system (6) we obtain;

$$\dot{s}(\underline{e},t) = \delta_s + \ddot{y}_d - f(\underline{x},t) - u(t) - D(\underline{x},t) \tag{8}$$

If $f(\underline{x},t)$ is known and free of external disturbancesanduncertainties, and when the system (6) is restricted to the $(\underline{e},t) = 0$, it will be governed by an equivalent control $u_{eq}$ obtained by:

$$u_{eq} = -\left[f(\underline{x},t) - \ddot{y}_d - \delta_s\right] \tag{9}$$

The global control is composed of the equivalent control and the Super Twisting terms $u_1$ and $u_2$ such that;

$$\begin{cases} \dot{u}_1 = -\lambda_1 sign(s(\underline{e},t)) \\ u_2 = -\lambda_2 |s(\underline{x},t)|^{(1/2)} sign(s(\underline{e},t)) \end{cases} \tag{10}$$

where $\lambda_1$ and $\lambda_2$, are the Super Twisting control gains [21], by adding these term to (9), we obtain the global control:

$$u = -\left[f(\underline{x},t) - \ddot{y}_d - \delta_s - \int_0^T \dot{u}_1 - u_2\right] \tag{11}$$

The sufficient condition to ensure the transition trajectory of the tracking error from approaching phase to the sliding one is:

$$\frac{1}{2}\frac{d}{dt}s^2(\underline{e},t) = s(\underline{e},t)\dot{s}(\underline{e},t) \le -\eta \left|s(\underline{e},t)\right| \tag{12}$$

where $\eta > 0$ is a constant.

After some manipulations, we obtain:

$$-\lambda_1 t - \lambda_2 \left|s(\underline{e},t)\right|^{(1/2)} + D(\underline{x},t)\,sign\big(s(\underline{e},t)\big) \le -\eta \qquad (13)$$

Then we can choose the parameters of $\lambda_1$ and $\lambda_2$ as follows:

$$\lambda_1 t + \lambda_2 \left|s(\underline{e},t)\right|^{(1/2)} \ge \eta + \left|D(\underline{x},t)\right| \\ \ge \eta + \Delta \qquad (14)$$

Note that the control law (11) depends only on the parameters $\lambda$, $\lambda_1$, $\lambda_2$, and nonlinear continuous function $f(\underline{x},t)$. However, the knowledge of the $D's$ upper bound and $f(\underline{x},t)$ is required in the optimal choice of $\lambda_1$ and $\lambda_2$, in the approaching phase. Therefore $f(\underline{x},t)$ is unknown and $(\underline{x},t) \ne 0$.

In the rest of paper we solved these problems by introducing an adaptive fuzzy second order sliding mode controller.

## 2.2. Interval Type-2 Fuzzy Logic System

Fuzzy Logic Systems (FLSs) are known as the universal approximators and have several applications in control design and identification. A type-1 fuzzy system consists of four major parts: fuzzifier, rule base, inference engine, and defuzzifier. A T2FLS is very similar to a T1FLS [27], the major structure difference being that the defuzzifier block of a T1FLS is replaced by the output processing block in a T2FLS, which consists of type-reduction followed by defuzzification.

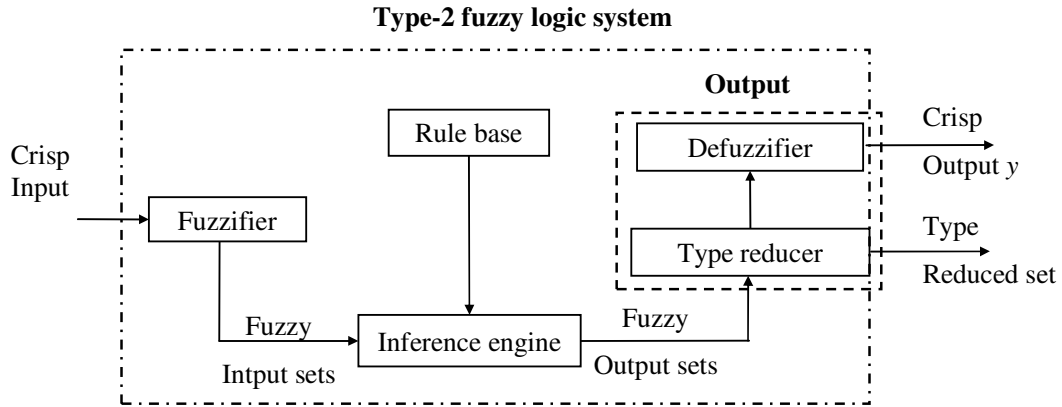

Figure 1. Structure of a type-2 fuzzy logic system.

In a T2FS, a Gaussian function with a known standard deviation is chosen, while the mean ($m$) varies between $m_1$ and $m_2$. Therefore, a uniform weighting is assumed to represent a footprint of uncertainty as shaded in Figure. 2.

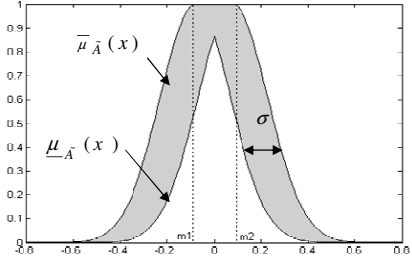

Figure 2. Interval type-2 Gaussian fuzzy set.

It is clear that the type-2 fuzzy set is in a region bounded by an upper MF and a lower MF denoted as $\bar{\mu}_{\tilde{A}}(x)$ and $\underline{\mu}_{\tilde{A}}(x)$ respectively, and is named a foot of uncertainty (FOU). Assume that there are $M$ rules in a type-2 fuzzy rule base, each of which has the following form:

$$R^i: \quad IF\, x_1\, is\, \tilde{F}_1^i, and\, ...\, , and\, x_n\, is\, \tilde{F}_n^i \, , THEN\, y\, is\, [w_l^i w_r^i]$$

where $x_j$, $j=1,2,...,n$ and $y$ are the input and output variables of the type-2 fuzzy system, respectively, the $\tilde{F}_n^i$ is the type-2 fuzzy sets of antecedent part, and $[w_l^i w_r^i]$ is the weighting interval set in the consequent part. The operation of type-reduction is to give a type-1 set from a type-2 set. In the meantime, the firing strength $F^i$ for the $ith$ rule can be an interval type-2 set expressed as;

$$F^i \equiv [\underline{f}^i, \overline{f}^i] \tag{15}$$

where

$$\begin{cases} \underline{f}^i = \underline{\mu}_{\tilde{F}_1^i}(x_1) * ... * \underline{\mu}_{\tilde{F}_n^i}(x_n) \\ \overline{f}^i = \bar{\mu}_{\tilde{F}_1^i}(x_1) * ... * \bar{\mu}_{\tilde{F}_n^i}(x_n) \end{cases} \tag{16}$$

In this paper, the center of set type-reduction method is used to simplify the notation. Therefore, the output can be expressed as;

$$y_{cos}(x) = [y_l, y_r]$$
$$= \int_{w^1 \in [w_l^1, w_r^1]} ... \int_{w^M \in [w_l^M, w_r^M]} \times \int_{f^1 \in [\underline{f}^1, \overline{f}^1]} ... \int_{f^M \in [\underline{f}^M, \overline{f}^M]} 1 \Big/ \frac{\sum_{i=1}^M f^i w^i}{\sum_{i=1}^M f^i} \tag{17}$$

where $y_{cos}(x)$ is also an interval type-1 set determined by left and right most points ($y_l$ and $y_r$), which can be derived from consequent centroid set $[w_r^i, w_l^i]$ (either $\underline{w}^i$ or $\overline{w}^i$) and the firing strength $f^i \in F^i = [\underline{f}^i, \overline{f}^i]$. The interval set $[w_r^i, w_l^i]$ ($i=1,... ,M$)should be computed or set first before the computation of $y_{cos}(x)$. For any value $y \in y_{cos}$. Hence, left-most point $y_l$ and right-most point $y_r$ can be expressed as [27];

$$y_l = \frac{\sum_{i=1}^M f_l^i w_l^i}{\sum_{i=1}^M f_l^i} \quad \text{and} \quad y_r = \frac{\sum_{i=1}^M f_r^i w_r^i}{\sum_{i=1}^M f_r^i} \tag{18}$$

Using the center of set type-reduction method to compute $y_l$ and $y_r$. Hence, $y_l$ and $y_r$ in (18) can be re-expressed as;

$$
\begin{aligned}
y_r &= y_r(\underline{f}^1, \dots, \underline{f}^R, \overline{f}^{R+1}, \dots, \overline{f}^M, w_r^1, \dots, w_r^M) \\
&= \left(\sum_{i=1}^R \underline{f}^i w_r^i + \sum_{i=R+1}^M \overline{f}^i w_r^i\right) \Big/ \left(\sum_{i=1}^R \underline{f}^i + \sum_{i=R+1}^M \overline{f}^i\right)
\end{aligned}
\tag{19}
$$

$$
\begin{aligned}
y_l &= y_l(\overline{f}^1, \dots, \overline{f}^L, \underline{f}^{L+1}, \dots, \underline{f}^M, w_l^1, \dots, w_l^M) \\
&= \left(\sum_{i=1}^L \overline{f}^i w_l^i + \sum_{i=L+1}^M \underline{f}^i w_l^i\right) \Big/ \left(\sum_{i=1}^L \overline{f}^i + \sum_{i=L+1}^M \underline{f}^i\right)
\end{aligned}
\tag{20}
$$

The defuzzified crisp output from an IT2FLS is the average of $y_l$ and $y_r$, that is:

$$
y(x) = \frac{y_l + y_r}{2}
\tag{21}
$$

# 3. Adaptive Interval Type-2 Fuzzy Second Order Sliding Mode Control

In this section, the unknown function $f(\underline{x}, t)$ and switching signals of the super twisting terms $u_1$ and $u_2$ will be replaced by adaptive type-2 fuzzy systems., then we replace $f(\underline{x}, t)$ and the Super Twisting terms by $\hat{f}(\underline{x}, \underline{\theta}_f), \hat{u}_1(s, \underline{\theta}_1)$ and $\hat{u}_2(s, \underline{\theta}_2)$ respectively such that:

$$
\hat{f}(\underline{x}, \underline{\theta}_f) = \underline{\theta}_f^T \underline{\xi}_f(\underline{x})
\tag{22}
$$

$$
\hat{u}_1(s, \underline{\theta}_1) = \underline{\theta}_1^T \underline{\xi}_1(s)t
\tag{23}
$$

$$
\hat{u}_2(s, \underline{\theta}_2) = \left|s(\underline{e}, t)\right|^{(1/2)} \underline{\theta}_2^T \underline{\xi}_2(s)
\tag{24}
$$

where $\underline{\theta}_f$, $\underline{\theta}_1$ and $\underline{\theta}_2$ are adjustable parameters vectors.

To guarantee the global stability of closed loop system (6) with the convergence of tracking errors to zero, we propose the following control law:

$$
u = -\left[\hat{f}(\underline{x}, \underline{\theta}_f) - \delta_s - \ddot{y}_d - \hat{u}_1(s, \underline{\theta}_1) - \hat{u}_2(s, \underline{\theta}_2)\right]
\tag{25}
$$

In order to derive the adaptive laws of adjusting $\underline{\theta}_f$, $\underline{\theta}_1$ and $\underline{\theta}_2$, first, we define the optimal parameter vector $\underline{\theta}_f^*$, $\underline{\theta}_1^*$ and $\underline{\theta}_2^*$ as;

$$
\underline{\theta}_f^* = \underset{\underline{\theta}_f \in \Omega_f}{\operatorname{argmin}}\left[\sup_{x \in \Omega_x}\left|\hat{f}(\underline{x}, \underline{\theta}_f) - f_0(\underline{x}, t)\right|\right], \underline{\theta}_1^* = \underset{\underline{\theta}_1 \in \Omega_1}{\operatorname{argmin}}\left[\sup_{s \in \Omega_s}\left|\hat{u}_1(s, \underline{\theta}_1) - u_1\right|\right]
$$

and $\underline{\theta}_2^* = \underset{\underline{\theta}_2 \in \Omega_2}{\operatorname{argmin}}\left[\sup_{s \in \Omega_s}\left|\hat{u}_2(s, \underline{\theta}_2) - u_2\right|\right].$

where $\Omega_f$, $\Omega_1$, $\Omega_2$, $\Omega_x$ and $\Omega_s$ are constraint sets of suitable bounds on $\theta_f$, $\theta_1$, $\theta_2$, $x$ and $s$, respectively, they are defined as;

$$\Omega_f = \left\{ \underline{\theta}_f : \left| \underline{\theta}_f \right| \leq M_f \right\}, \Omega_1 = \left\{ \underline{\theta}_1 : \left| \underline{\theta}_1 \right| \leq M_1 \right\}, \Omega_2 = \left\{ \underline{\theta}_2 : \left| \underline{\theta}_2 \right| \leq M_2 \right\},$$
$$\Omega_x = \{ x : |x| \leq M_x \}, \quad \Omega_s = \{ s : |s| \leq M_s \};$$

Where $M_f, M_1, M_2, M_x$ and $M_s$ are positive constants.

The minimum approximation error is defined as;

$$w = \left[ f(\underline{x}, \mathrm{t}) - \hat{f}(\underline{x}, \underline{\theta}_f^*) \right]$$

We can write,

$$|w| \leq \left| f(\underline{x}, \mathrm{t}) - \hat{f}(\underline{x}, \underline{\theta}_f^*) \right|$$
$$\leq \left| f(\underline{x}, \mathrm{t}) \right| - \left\| \underline{\theta}_f^{*T} \right\| \left\| \underline{\xi}_f(\underline{x}) \right\| \leq F - M_f$$

By using $F - M_f = \beta$, it can be easily concluded that $w$ is bounded $w \leq \beta$,

Then the optimal parameters of $f(\underline{x}, \mathrm{t})$, $u_1$ and $u_2$ are defined as:

$$\hat{f}(\underline{x}, \underline{\theta}_f^*) = \underline{\theta}_f^{*T} \underline{\xi}_f(\underline{x}) \tag{26}$$

$$\hat{u}^*_1(s, \underline{\theta}_1^*) = \underline{\theta}_1^{*T} \underline{\xi}_1(s) t \tag{27}$$

$$\hat{u}^*_2(s, \underline{\theta}_2^*) = \left| s(\underline{e}, t) \right|^{(1/2)} \underline{\theta}_2^{*T} \underline{\xi}_2(s) \tag{28}$$

From the study of the closed loop stability, we can find the adaptation laws of adjustable parameters, then, we consider the following Lyapunov function:

$$V = \frac{1}{2} s^2 + \frac{1}{2\gamma_f} \underline{\tilde{\theta}}_f^{\ T} \underline{\tilde{\theta}}_f + \frac{1}{2\gamma_1} \underline{\tilde{\theta}}_1^{\ T} \underline{\tilde{\theta}}_1 + \frac{1}{2\gamma_2} \underline{\tilde{\theta}}_2^{\ T} \underline{\tilde{\theta}}_2 \tag{29}$$

where $\underline{\tilde{\theta}}_i = \underline{\theta}_i - \underline{\theta}_i^*$, $(i = 1,2)$ and $\underline{\tilde{\theta}}_f = \underline{\theta}_f - \underline{\theta}_f^*$. $\gamma_f$, $\gamma_1$ and $\gamma_2$ are positive training constants, the time derivative of (29) is :

$$\dot{V} = \dot{s}(\underline{e}, t) s(\underline{e}, t) + \frac{1}{\gamma_f} \underline{\tilde{\theta}}_f^{\ T} \underline{\dot{\theta}}_f + \frac{1}{\gamma_1} \underline{\tilde{\theta}}_1^{\ T} \underline{\dot{\theta}}_1 + \frac{1}{\gamma_2} \underline{\tilde{\theta}}_2^{\ T} \underline{\dot{\theta}}_2 \tag{30}$$

By using the control law (25), the equation (22-24), the time derivative of the sliding surface (8) becomes:

$$\dot{s} = f(\underline{x}, t) + D(\underline{x}, t) - \hat{f}(\underline{x}, \underline{\theta}_f^*) + \hat{u}_1(s) + \hat{u}_2(s)$$
$$= f(\underline{x}, t) - \hat{f}(\underline{x}, \underline{\theta}_f) + \hat{f}(\underline{x}, \underline{\theta}_f^*) - \hat{f}(\underline{x}, \underline{\theta}_f^*) + D(\underline{x}, t) + \hat{u}_1(s) + \hat{u}_1^*(s) - \hat{u}_1^*(s) + \hat{u}_2(s) + \hat{u}_2^*(s) - \hat{u}_2^*(s)$$
$$= w - (\underline{\theta}_f - \underline{\theta}_f^*)^T \underline{\xi}_f(\underline{x}) + (\underline{\theta}_1 - \underline{\theta}_1^*)^T \underline{\xi}_1(s) t + \hat{u}_1^*(s) + \hat{u}_2^*(s) + (\underline{\theta}_2 - \underline{\theta}_2^*)^T |s(\underline{e}, t)|^{(1/2)} \underline{\xi}_2(s) + D(\underline{x}, t)$$
$$\tag{31}$$

The substitution of (31) in (30) gives:

$$\dot{V} = \frac{1}{\gamma_f}\underline{\tilde{\theta}}_f \ (\underline{\dot{\theta}}_f \ - \gamma_f s(\underline{e},t)\underline{\xi}_f(\underline{x})) + s(\underline{e},t)(w + \hat{u}_1^*(s) + \hat{u}_2^*(s) + D(\underline{x},t))$$
$$+ \frac{1}{\gamma_1}\underline{\tilde{\theta}}_1 \ (\underline{\dot{\theta}}_1 \ + \gamma_1 s(\underline{e},t)\underline{\xi}_1(s)t) \ + \frac{1}{\gamma_2}\underline{\tilde{\theta}}_2 \ (\underline{\dot{\theta}}_2 \ + \gamma_2 s(\underline{e},t)\big|s(\underline{e},t)\big|^{(1/2)}\underline{\xi}_2(s))$$

(32)

By choosing the following adaptation laws:

$$\underline{\dot{\theta}}_f \ = \gamma_f s(\underline{e},t)\underline{\xi}_f(\underline{x})$$

(33)

$$\underline{\dot{\theta}}_1 \ = -\gamma_1 s(\underline{e},t)\underline{\xi}_1(s)t$$

(34)

$$\underline{\dot{\theta}}_2 \ = -\gamma_2 s(\underline{e},t)\big|s(\underline{e},t)\big|^{(1/2)}\underline{\xi}_2(s)$$

(35)

where $\underline{\dot{\tilde{\theta}}}_i \ = \underline{\dot{\theta}}_i \ ,(i=1,2)$ and $\underline{\dot{\tilde{\theta}}}_f \ = \underline{\dot{\theta}}_f$ . Therefore, we obtain:

$$\dot{V} = s(w + D(\underline{x},t) - \lambda_1^* t sign(s) - \lambda_2^*|s|^{(1/2)} sign(s))$$

(36)

$$\dot{V} = ws + D(\underline{x},t) \ s - (\lambda_1^* t + \lambda_2^*|s|^{(1/2)})|s|$$
$$\leq -\eta|s| + |w||s| \leq -\eta + \beta$$

(37)

According to Barbalat's lemma [28], we can state that the sliding surface is constructed to be attractive and $\lim_{t\to\infty} e(t) = 0$. Therefore, the control objective is achieved, and hence, we can synthesize the robust controller based on second order sliding mode and fuzzy type-2 systems, in which we can force the output system $\underline{x}$ to follow a bounded reference trajectory $\underline{y}_d$.

The overall scheme of the adaptive type-2 fuzzy second order sliding mode control for nonlinear chaotic system in presence of uncertainties, external disturbance and the training data is corrupted with internal noiseis shown in Figure. 3.

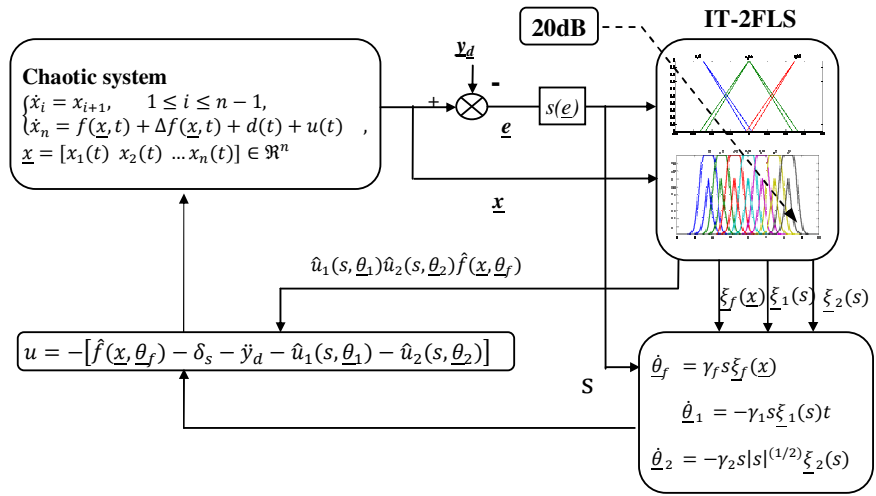

Figure 3. Overall adaptive type-2 fuzzy second order sliding mode control scheme in presence of noise.

## 4. SIMULATION EXAMPLE

The above described control scheme is now used to stabilize the nonlinear chaotic system which is defined as follows;

$$\begin{cases} \dot{x}_1 = x_2, \\ \dot{x}_2 = -0.4x_2 - 1.1x_1 - x_1{}^3 - 2.1\cos(1.8t) \end{cases} \tag{38}$$

With initial states$(0) = [0.1 \ 0]^T$.

For free input, the simulation results of system are shown in Figure 4-5.

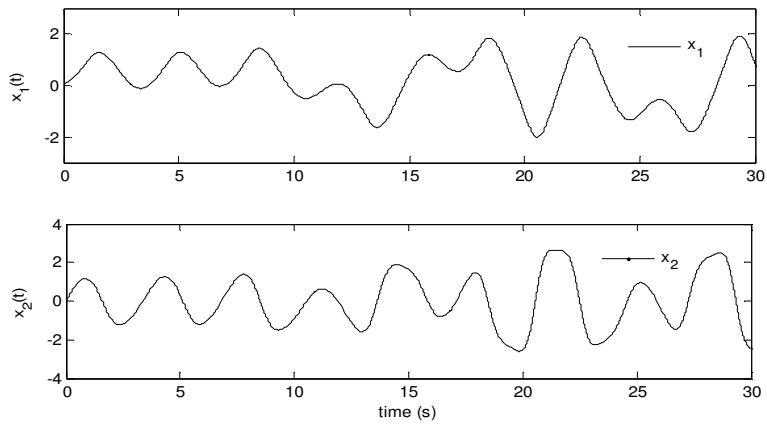

Figure 4. Time response of states ($x_1$, $x_2$)

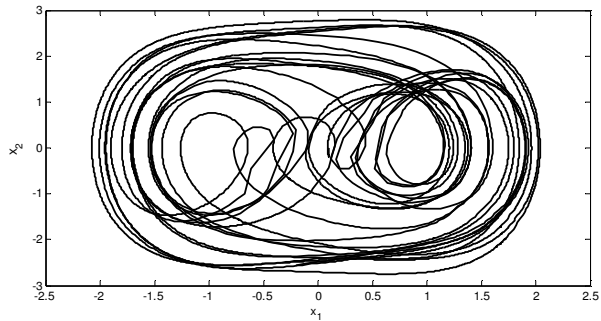

Figure 5. Typical chaotic behavior of duffingoscillator

The control objective is to force the states system $x_i(t), i = 1,2$ to track the reference trajectories $y_d(t)$ and $\dot{y}_d(t)$ in finite time, such as $y_d(t) = (\pi/3)(\sin(t) + 0.3\sin(3t))$, the adaptive interval type-2 fuzzy second order sliding mode control (25) is added into the system as follows:

$$\begin{cases} \dot{x}_1 = x_2, \\ \dot{x}_2 = -0.4x_2 - 1.1x_1 - x_1{}^3 - 2.1\cos(1.8t) + \Delta f(x,t) + d(t) + u(t) \end{cases} \tag{39}$$

The sliding surface is selected as: $s = \dot{e} + \lambda e$; where $\lambda = 10$, and the adaptive parameters $\gamma_1 = 10$, $\gamma_2 = 6$ and $\gamma_f = 15$. To designthe equivalent part of control signal, the input variables of the fuzzy system $\hat{f}(\underline{x}, \theta_f)$are chosen as $x_i(t), i = 1,2$, and we define seven type-2 Gaussian membership functions selected as $F_i^l, l = 1, \ldots, 7$which are shown in table. 1, with variance $\sigma = 0.5$ and initial values $\theta_f(0) = O_{2 \times 7}$.

Similarly to generate the two adaptive fuzzy systems which allow us to approximate the reaching part of control signal ($u_1$and $u_2$), we consider three type-2 fuzzy interval sets according to the variable $s(t)$ (Figure. 6).

Table 1. Interval Type-2 Fuzzy Membership Functions For $x_i (i = 1,2)$.

| | **Mean** | | | **Mean** | |
|---|---|---|---|---|---|
| | $m_1$ | $m_2$ | | $m_1$ | $m_2$ |
| $\mu_{F_i^1}(x_i)$ | -3.5 | -2.5 | $\mu_{F_i^5}(x_i)$ | 0.5 | 1.5 |
| $\mu_{F_i^2}(x_i)$ | -2.5 | -1.5 | $\mu_{F_i^6}(x_i)$ | 1.5 | 2.5 |
| $\mu_{F_i^3}(x_i)$ | -1.5 | -0.5 | $\mu_{F_i^7}(x_i)$ | 2.5 | 3.5 |
| $\mu_{F_i^4}(x_i)$ | -0.5 | 0.5 | | | |

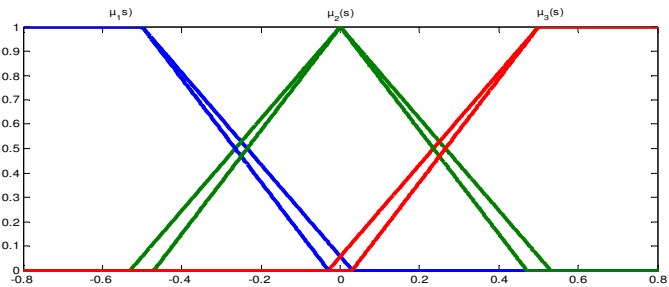

Figure6. Interval type-2 antecedent membership functions of $s(t)$

The simulation results are presented in the presence of uncertainties$\Delta f(\underline{x}, t) = (\pi/6)\sin(2\pi x_1(t))\sin(3\pi x_2(t))$, external disturbance$d(t) = \sin(2t)$, and white Gaussian noise is applied to the measured signal $x_i(t)$, $i = 1,2$and $s$with Signal to Noise Ratios (SNR=20dB), with initial states $x(0) = [1 \ 0]^T$

The tracking performance of states $\underline{x}(t)$is shown in Figures7-8. The tracking errors and control input $u(t)$are shown in Figures 9-10, the phase-plane trajectories of system are represented infigures 11-12, and the sliding manifold with its time derivative in figure 13.

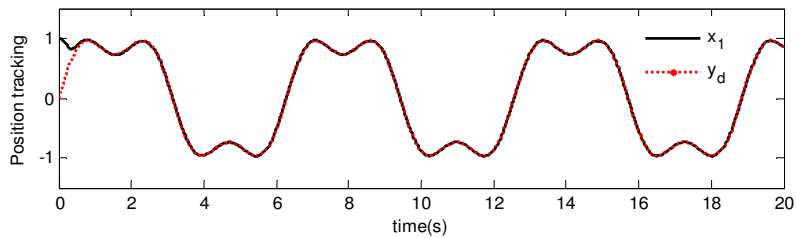

Figure 7. Time response of state $x_1$ and desired trajectory $x_d$.

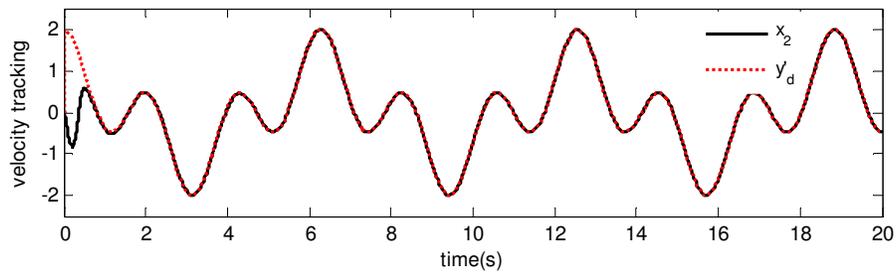

Figure 8. Time response of state $x_2$ and desired trajectory $\dot{x}_d$

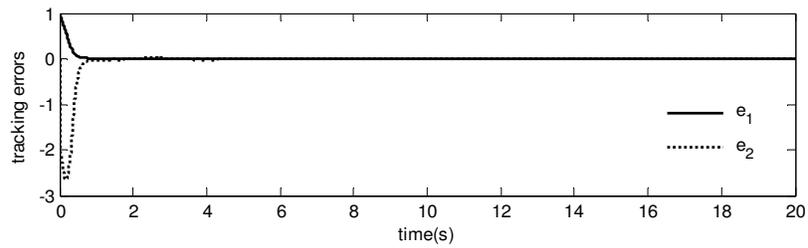

Figure 9. Tracking errors $e_1(t)$ and $e_2(t)$

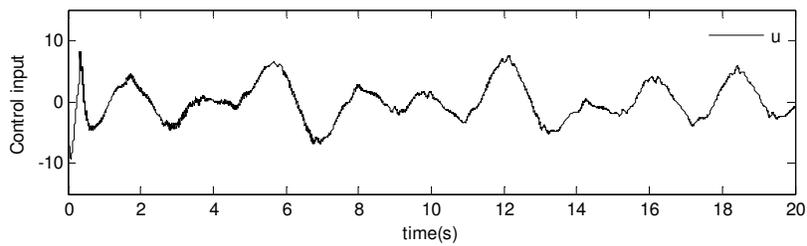

Figure 10. Control input $u(t)$

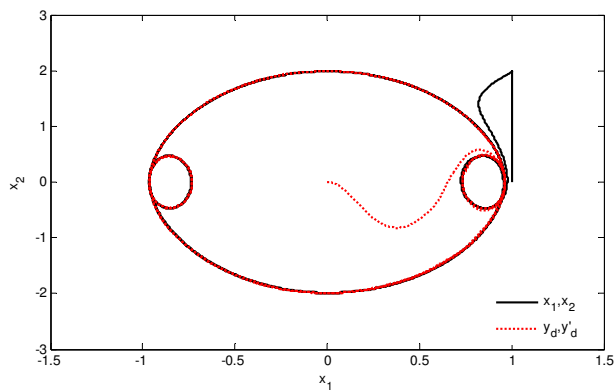

Figure 11.System state space of duffing oscillator

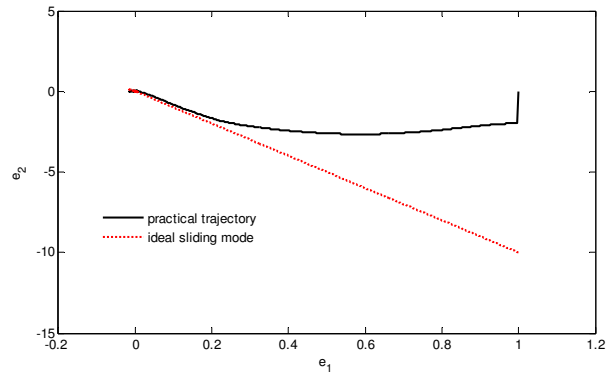

Figure 12.Phase-plane trajectory of tracking errors*(e₁,e₂)*

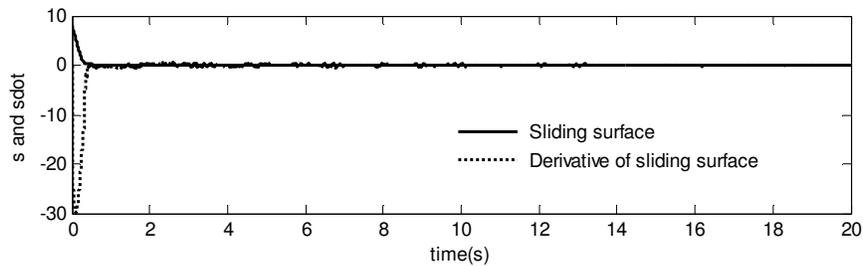

Figure 13. Trajectories of sliding manifold *s* and its derivative *ṡ*

According to the above simulation results, it is obvious that the tracking errors converge to zero in a finite time, which implies that the proposed controller forces the system states to reach quickly their references. Obviously, the phase trajectory of ($e_1$, $e_2$) converges directly to the phase origin. In the same time, the implementation of Super Twisting algorithm in higher order sliding mode control allows obtaining a smooth control signal (Figure10).

## 5. CONCLUSION

In this paper, the problem of stabilization orbit of uncertain chaotic system working in the presence of uncertainties, external and internal disturbances is solved by incorporation of adaptive interval type-2 control scheme and second order sliding mode approach using super-twisting algorithm. The adaptive interval type-2 fuzzy systemsare introduced to approximate the unknown part of system and Super Twisting gains. Based on the Laypunov stability criterion, the adaptation law of adjustable parameters of the type-2 fuzzy system and the stability of closed loop system are ensured. A simulation example has been presented to illustrate the robustnessand the effectiveness of the proposed approach.

## Authors

**Rim Hendel** received here engineering and Master degrees in Automatic from Setif University (Setif 1), Algeria, in 2009 and 2012 respectively. From November 2012, she is Ph.D. student in the Engineering Faculty with the QUERE laboratory at the University ofSetif 1. Here research interests are higher order sliding mode control, fuzzy type-1 and type-2 systems, nonlinear systems 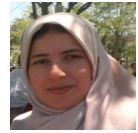

**Faridkhaber** received his D.E.A in 1990 and his Master in 1992 degrees in industrial control, and his PhD in 2006 from Setif University (Setif 1), Algeria, in automatic control. He is currently a Professor in the Engineering Faculty from the same university. His research interests include multivariable adaptive control, LMI control and type-2 fuzzy control of renewable energy systems 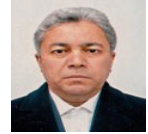

**NajibEssounbouli** received his Maitrise from the University of Sciences and Technology of Marrakech (FSTG) in Morocco, his D.E.A. in 2000, his Ph.D. in 2004, and its Habilitation from Reims University of Champagne- Ardenne, all in Electrical Engineering. From September 2005 to 2010, he has been an Assistant Professor with IUT of Troyes, Reims Champagne Ardenne University. He is a currently a Professor and Head 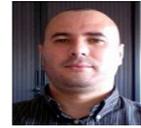 of the Mechanical Engineering Department of IUT at Troyes, Reims University. His current research interests are in the areas of fuzzy logic control, robust adaptive control, renewable energy and control drive.